\documentstyle[prl,aps,twocolumn,floats]{revtex}
\begin{document} 

\title{Constrained randomization of time series data}
\date{Phys.\ Rev.\ Lett.\ {\bf 80} (1998) 2105}
\author{Thomas Schreiber\\
   Physics Department, University of Wuppertal,
   D-42097 Wuppertal, Germany}
\maketitle

\begin{abstract} 
A new method is introduced to create artificial time sequences that fulfil
given constraints but are random otherwise. Constraints are usually derived
from a measured signal for which surrogate data are to be generated. They are
fulfilled by minimizing a suitable cost function using simulated annealing.
A wide variety of structures can be imposed on the surrogate series, including
multivariate, nonlinear, and nonstationary properties. When the linear
correlation structure is to be preserved, the new approach avoids certain
artifacts generated by Fourier-based randomization schemes. PACS: 05.45.+b
\end{abstract}

Randomization of data and Monte Carlo resampling of probability distributions
is a common technique in statistics~\cite{efron}. In the context of nonlinear
time series analysis it has been discussed by several authors and is usually
referred to as the {\em method of surrogate data}~\cite{surro}. A null
hypothesis for the nature of a time series can be tested by comparing the
value of an observable $\gamma$ obtained using the data with values obtained
using a collection of {\em surrogate} time series representing the null
hypothesis.  All but the simplest null assumptions allow for certain
structures, for example linear serial correlations. There are two distinct
ways to implement such structures when creating surrogate series. Traditional
bootstrap methods use explicit model equations that have to be extracted from
the data.  This {\em typical realizations} approach can be very powerful for
the computation of confidence intervals, provided the model equations can be
extracted successfully. As discussed in Ref.~\cite{tp}, the alternative
approach of {\em constrained realizations} is more suitable for the purpose of
hypothesis testing. It avoids the fitting of model equations by directly
imposing the desired structures onto the randomized time series. However, the
choice of possible null hypothesis has so far been limited by the difficulty
of imposing arbitrary structures on otherwise random sequences. Algorithms
exist mainly for the following cases. (1) The null hypothesis of independent
random numbers from a fixed but unknown distribution can be tested against
permutations without repetition of the data since these conserve the sample
distribution exactly. (2) The case of Gaussian noise with arbitrary linear
correlations leads to the Fourier transform method. The Fourier transform of
the data is multiplied by random phases and then transformed back, conserving
the sample periodogram. (See Ref.~\cite{multi} for the multivariate case.)
(3) Surrogates with a given distribution and given linear correlations are
needed for the null hypothesis of a monotonically rescaled Gaussian linear
stochastic process. This is approximately achieved by the amplitude adjusted
Fourier transform (AAFT) algorithm~\cite{surro} and the more accurate
iterative method proposed in Ref.~\cite{we}.

This paper will introduce a general method for generating random time
sequences subject to quite general constraints. Any null hypothesis that leads
to a complete set of observables can thus be tested for. All the above cases
can be dealt with (often with higher accuracy), but also multivariate,
nonstationary, nonlinear or other constraints can be implemented.  In all the
applications in this paper, the single time probability distribution will be
one of the constraints, leading to the requirement that the randomized
sequence is a permutation of a fixed collection of values. All other
constraints, for example part or all of the lags of the autocorrelation
function, will be formulated in terms of a cost function which is then
minimized among all possible permutations by the method of simulated
annealing.

After giving the actual randomization scheme I will discuss some major
applications. We will show that the algorithm yields a more accurate
nonlinearity test and avoids known artifacts that are introduced by end effects
with ordinary, Fourier-based surrogates~\cite{sfi}. We will also give examples
with more general null hypothesis than that of a rescaled stationary linear
stochastic process. For these examples, previous methods could not provide
appropriate surrogates.

The algorithm is conceptually very simple:
\begin{enumerate} 
\item Specify constraints ${\cal C}_i(\{{\tilde x}_n\})=0$ in terms of a 
   {\em cost function} $E(\{{\tilde x}_n\})$, constructed
  to have a global minimum when the constraint is fulfilled. 
\item Minimize $E(\{{\tilde x}_n\})$ among all permutations $\{{\tilde x}_n\}$ 
   of a time series $\{x_n\}$ by simulated annealing. Configurations
   are updated by exchanging pairs in $\{{\tilde x}_n\}$.
\end{enumerate}
Examples of its use will be given below.

The simulated annealing method is particularly useful for combinatorial
minimization with false minima. It goes back to Metropolis {\em et
  al.}~\cite{metro}, and is thoroughly discussed in the
literature~\cite{anneal}. 
Essentially, the cost function is interpreted as an energy in a thermodynamic
system. At some finite ``temperature'' $T$, system configurations are visited
consecutively with a probability according to the Boltzmann distribution
$e^{-E/T}$ of the canonical ensemble. This is achieved by accepting changes of
the configuration with a probability $p=1$ if the energy is decreased $(\Delta
E<0)$ and $p=e^{-\Delta E/T}$ if the energy is increased, $(\Delta E\ge
0)$. The temperature is decreased slowly, thereby ``annealing'' the system to
the ground state of minimal ``energy'', that is, the minimum of the cost
function. In the limit $T\to 0$, all ground state configurations can be reached
with equal probability.  Although some general rigorous convergence results are
available, in practical applications of simulated annealing some
problem-specific choices have to be made. In particular, apart from the cost
function itself, one has to specify a method of updating the configurations and
a schedule for lowering the temperature. A way to efficiently reach all
permutations by small individual changes is by exchanging randomly chosen (not
necessarily close-by) pairs. In many cases, an exchange of two points is
reflected in a rather simple update of the cost function. This is important for
speed of computation.  Many cooling schemes have been discussed in the
literature~\cite{anneal}. In this work, the temperature is multiplied by
$\alpha$ at each cooling step. Cooling is done if either the number of
successful updates since the last cooling exceeds $N_{\rm succ}$, or the total
number of configurations visited during this cooling step exceeds $N_{\rm
total}$. It is difficult to give general rules on how to choose $\alpha, N_{\rm
succ},$ and $N_{\rm total}$.  Slow cooling is necessary if the desired accuracy
of the constraint is high.  It seems reasonable to increase $N_{\rm succ}$ and
$N_{\rm total}$ with the system size, but also with the number of constraints
incorporated in the cost function. Generally, one can choose a tolerance for
the constraints, start with rather fast cooling and repeat the analysis with a
slower cooling rate if the accuracy has not been met. Other more sophisticated
cooling schemes may be suitable depending on the specific situation. The reader
is referred to the standard literature~\cite{anneal}.

Let us first demonstrate that the algorithm yields more accurate results than
previous methods for the most prominent application of surrogate data, which is
statistical testing for nonlinearity in a time series. Consider the null
hypothesis that there is a sequence $\{y_n\}$ that has been generated by a
Gaussian linear stochastic process. As the only allowed kind of nonlinearity,
the actual data $\{x_n\}$ consists of observations of $\{y_n\}$ made through a
monotone instantaneous measurement function: $x_n=f(y_n)$. As
discussed e.g. in Ref.~\cite{we}, the corresponding Monte Carlo sample has to
be constrained to have (i) the same single time probability distribution and
(ii) the same sample auto-covariance function~\cite{footC}
\begin{equation}\label{eq:C}
    C(\tau)={1\over N-\tau}\sum_{n=\tau}^{N-1} x_n x_{n-\tau}
\end{equation}
for all lags $\tau=0,\ldots,N-1$. (Zero mean has been imposed for simplicity of
notation.)

In the actual test, a nonlinear observable $\gamma$ is computed for the data
and a collection of surrogate data sets. (See Ref.~\cite{ss} for a comparison
of the performance of different statistics $\gamma$.) The null hypothesis will
be rejected if the result $\gamma_0$ obtained for the data is incompatible with
the probability distribution of $\gamma$ estimated from the surrogates. Note
that although even different realizations of the same process will have the
same sample auto-covariance function only up to statistical fluctuations, it is
essential that the surrogates are constrained to $C(\tau)^{\rm (data)}$ as
accurately as possible--since almost every discriminating statistic $\gamma$
will depend on $C(\tau)$, we are otherwise likely to introduce a bias and
possibly spurious rejections of the null hypothesis. See also the discussion in
Ref.~\cite{tp}.

Previous attempts to implement the above constraints have only been partially
successful.  In the scheme introduced here, property (i) is easily implemented
by considering as candidates for randomized series all permutations of the
measured time sequence $\{y_n\}$. Requirement (ii) can be achieved by finding a
permutation of $\{y_n\}$ which, within the desired accuracy, minimizes a cost
function like the following~\cite{footCost}:
\begin{equation}
    E^{(q)}=\left[
               \sum_{\tau=0}^{N-1} |C(\tau)-C^{\rm (data)}(\tau)|^q
            \right]^{1/q}
\,.\end{equation}
Provided the annealing scheme is brought to convergence with high accuracy,
the known artifacts that remain with previous approaches can be avoided.

\begin{table}[t]
   \begin{tabular}{llrl} 
   algorithm    & $\alpha$      & CPU time      & $(N/2)^{-1}E_p^{(\infty)}$\\
   \hline
   scramble     &               & ---           & 0.82 $\pm$ 0.02\\
   AAFT         &               & 0.01s         & 0.08 $\pm$ 0.02\\
   iterative    &               & 2s            & 0.03 $\pm$ 0.01\\
   annealing    & 0.8           & 2m            & 0.0055\\
                & 0.9           & 25m           & 0.0009\\
                & 0.98          & 10h           & 0.0003\\
   \end{tabular}
   \caption[]{\label{tab:1}Residual deviation from the desired 
   au\-to-co\-variance function for different methods of randomizing time
   series.} 
\end{table}

As discussed in~\cite{we}, the original (AAFT) algorithm~\cite{surro} can show
a bias towards a flat spectrum for short sequences. The iterative scheme
proposed in~\cite{we} removes this bias to a satisfactory approximation for
practical work.  Let us however compare the accuracy of the previously proposed
schemes to the present algorithm.  For comparability, a cost function is chosen
with respect to the time periodic sample auto-covariance function
\begin{equation}\label{eq:cp}
    C_p(\tau)={1\over N}
              \sum_{n=0}^{N-1} x_n x_{(n-\tau)\,{\rm\scriptsize mod}\,N}
\,.\end{equation}
which corresponds to the Fourier spectrum through the Wiener-Khinchin
theorem. Minimizing $E_p^{(\infty)} = \max_{\tau=0}^{N/2} |C_p(\tau)-C_p^{\rm
(data)}(\tau)|$ will reproduce the auto-covariance $C_p^{\rm (data)}$ measured
on the data. Time series of length $N=1000$ are generated by an autoregressive
model but measured using a nonlinear measurement function: $x_n=y_n^3,\quad
y_n=0.9 y_{n-1}+\eta_n$.  The residual maximal deviations of the
auto-covariances of the time series and surrogate sets were determined for (i)
random permutations of the data, (ii) usual AAFT surrogates~\cite{surro}, (iii)
surrogates created with the iterative scheme given in Ref.~\cite{we} and (iv)
outcomes of the annealing procedure for different cooling protocols. Note that
with slower cooling, arbitrarily high accuracy can be reached in principle.
Averages over 20 realizations were determined for cases (i) to (iii). The
iterative scheme (iii) was repeated until a fixed point was reached which was
the case after about 200 iterations.  Table~\ref{tab:1} summarizes the
results. Computation time time on a DEC alpha workstation at 400MHz clock rate
are given only for relative comparison.  The price for the superior accuracy of
the annealing scheme is its much higher computational cost.

As mentioned earlier, all previous randomization schemes~\cite{surro,we} make
use of the Fourier transform in order to achieve the desired linear correlation
structure. Note, however, that two sequences with the same Fourier amplitudes
do not quite have the same auto-covariance function $C(\tau)$,
eq.~(\ref{eq:C}). The Wiener-Khinchin theorem only says that the {\em periodic}
sample auto-covariance function $C_p(\tau)$, eq.~(\ref{eq:cp}), will be the
same. This amounts to assuming that the measured time series is exactly one
period of an infinite periodic signal, which is of course not what we believe
to be the case. The artifact generated by this flaw of previous algorithms has
been discussed e.g. in Ref.~\cite{sfi}. The periodically extended sequence may
undergo a phase slip or even a finite jump at $n=N$. The surrogate series will
have the power contained in that slip spread out over the whole observation
time, leading to additional high frequency content. Although spurious results
can be partially suppressed by selecting a segment of the data that
approximately returns to the initial value, it is desirable to preserve the
auto-covariance function $C(\tau)$ in eq.~(\ref{eq:C}) rather than $C_p(\tau)$
in eq.~(\ref{eq:cp}). With the annealing scheme proposed in this paper, this
can be easily done by choosing an appropriate cost function.

As an illustration, consider a particular autoregressive process of order two,
$x_n=1.3x_{n-1} -0.31x_{n-2} + \eta_n$. Since it is almost unstable, short
realizations often show a large difference between the first and the last
point. Periodic continuation turns this difference into a large step with broad
frequency content. For a realization of 160 points we found that for a
Fourier-based surrogate (method in Ref.~\cite{we}, same {\em periodic}
auto-covariance function $C_p(\tau)$), the sample autocorrelation $C(1)/C(0)$
was reduced from 0.92 to 0.85. Consequently, the power in the first differences
is increased by a factor of two and short term predictability is strongly
reduced. This can lead to spurious rejections of the null hypothesis of a
linear process. A sequence obtained by minimizing
$E^{(\infty)}=\max_{\tau=0}^{N-1} |C(\tau)-C^{\rm (data)}(\tau)|/\tau$ yielded
the correct value of $C(1)$ within $2\times 10^{-4}$.

\begin{figure}[t]
   \centerline{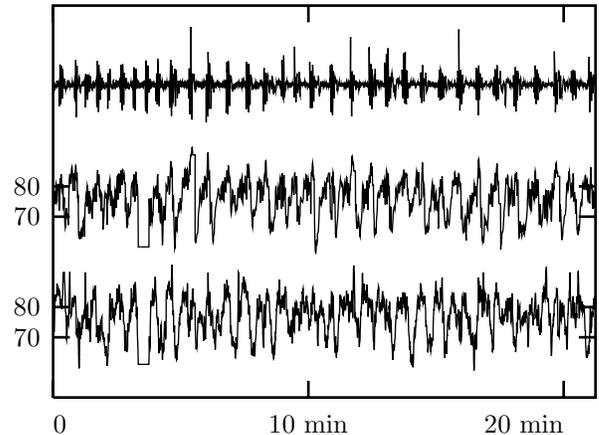}\vspace{10pt} 
   \caption[]{\label{fig:multi} Simultaneous measurements of breath and heart
   rates~\cite{gold}, upper and middle traces. Lower trace: a surrogate heart
   rate series preserving the autocorrelation structure and the
   cross-correlation to the fixed breath rate series, as well as a gap in the
   data. Auto- and cross-correlation together seems to explain some, but not
   all of the structure present in the heart rate series.}
\end{figure}

Apart from its potential for greater accuracy, the most striking feature of
the new scheme is its generality and flexibility. This point will be
demonstrated in the following examples which are by no means exhaustive. Note
that none of the examples below could be studied with previous surrogate data
schemes. Let us first study a multivariate example, a simultaneous recording of
the breath rate and the instantaneous heart rate of a human subject during
sleep. (Data set B of the Santa Fe Institute time series contest in
1991~\cite{gold}, samples 1800--4350.) Regarding the heart rate recording on
its own, one easily detects nonlinearity, in particular an asymmetry under time
reversal. An interesting question however is, how much of this structure can be
explained by linear dependence on the breath rate, the breath rate also being
non-time-reversible. In order to answer this question, one has to make
surrogates that have the same autocorrelation structure but also the same
cross-correlation with respect to the fixed input signal, the breath
rate. (Here the breath rate data is not randomized, which is of course also
possible within this framework.) Accordingly, a constraint is formulated
involving all lags up to 500 of the auto-covariance and the cross-covariance
($C_{xy}$) functions. The cost function is taken to be $\max_{\tau=0}^{500}
|C(\tau)-C^{\rm (data)}(\tau)|/\tau + \max_{\tau=-500}^{500} |C_{xy}(\tau) -
C_{xy}^{\rm (data)}(\tau)|/(|\tau|+1)$, other choices are possible. Further
suppose that during one minute the equipment spuriously recorded a constant
value. In order not to interpret this artifact as structure, the same artifact
is generated in the surrogates, simply by excluding these data points from the
permutation scheme.

\begin{figure}[t]
   \centerline{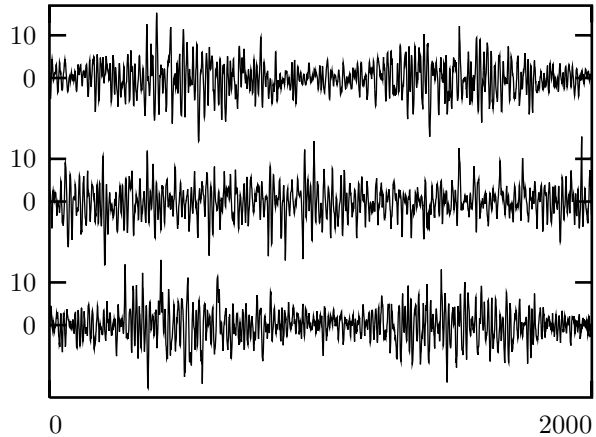}\vspace{10pt} 
   \caption[]{\label{fig:nonstat} A realization of a linear process with time
   dependent variance (upper), a usual AAFT surrogate (middle), and a
   surrogates with the same autocorrelations and the same running variance as
   the original series.}
\end{figure}

Figure~\ref{fig:multi} shows the measured breath rate (upper trace) and
instantaneous heart rate (middle trace). The lower trace shows a surrogate
conserving both, auto- and cross-correlations. The cooling rate was
$\alpha=0.95$, $N_{\rm succ}=10000$, $N_{\rm total}=3\times 10^5$. None of the
auto- and cross-covariances differed from the goal by more than $5\times
10^{-4}$ in units of the variance of the data after 3h of annealing.  (DEC
alpha workstation at 400MHz clock rate.) The visual impression from
Fig.~\ref{fig:multi} is that while the linear cross-correlation with the breath
rate explains the cyclic structure of the heart rate data, other features
remain unexplained. In particular, the surrogates don't show the asymmetry
under time seen in the data. Possible explanations of the remaining structure
include artifacts due to the peculiar way of deriving heart rate from
inter-beat intervals, nonlinear coupling to the breath activity, nonlinearity
in the cardiac system, and others.

Let us finally give a nonstationary example, an AR(2) process with periodically
modulated variance: $x_n=1.6x_{n-1} -0.8x_{n-2} + b_n\eta_n$ with $b_n=1+\sin^2
2\pi n/1000$. In Fig.~\ref{fig:nonstat} a realization ($N=2000$) is shown
together with two surrogate series. The first (middle trace) has been generated
by the AAFT algorithm, the second (lower trace) has been generated by the
annealing scheme to preserve the first 100 lags of $C(\tau)$ but also the
running variance in blocks of length 200, overlapping by 100.

In this paper it has been demonstrated that randomization under a wide variety
of constraints can be achieved with a permutation scheme that minimizes a
suitable cost function using simulated annealing. The approach is very general.
Constraints are not restricted to linear correlations. Multivariate, nonlinear,
but also time dependent, nonstationary properties can be easily implemented. A
wider range of examples will be studied elsewhere~\cite{malsehen}.
 
Resampling with constraints is the method of choice for hypothesis testing,
where it is preferable to parametric bootstrap methods. Although a general,
nonparametric resampling scheme has been introduced in this paper, care has to
be taken when similar ideas are to be exploited for the determination of error
bounds. The variance of statistical estimators usually depends on the
constraints imposed. To which extent reliable error distributions can be
obtained by selecting a minimal set of constraints and using resampling with
replacement will be a subject of future work.

I thank James Theiler, Daniel Kaplan, Thomas Sch\"urmann, Holger Kantz, Rainer
Hegger, and Eckehard Olbrich for stimulating discussions, and the Max Planck
Institute for Physics of Complex Systems in Dresden for kind
hospitality. This work was supported by the SFB 237 of the Deutsche
Forschungsgemeinschaft.


\begin{thebibliography}{10}
\bibitem{footC}
   Strictly speaking, the two constraints over-specify the problem. There are
   at most finitely many solutions $\{x'_n\}$, including the original sequence
   and the time reversed sequence $\{x_{N-n-1}\}$.  However, We will not
   usually find the absolute minimum, or otherwise we could penalize similarity
   to the original sequence.
\bibitem{footCost}
   In principle, any function that has a global minimum at $C(\tau)\equiv
   C^{\rm (data)}(\tau)$ can be used. In most cases we found good
   convergence using $E^{(q)}$ at $q=1$, $q=2$, or $q=\infty$. ($E^{(\infty)}$
   simply picks the maximal deviation.) For highly correlated sequences we
   found it advantageous to enhance the convergence at small lags by weighting
   the terms in the cost function by $1/\tau$. Some structures are rather
   difficult to build up by interchanging arbitrary pairs. Different weights
   in the cost function can help, but also alternative updating schemes are
   possible.
\end{thebibliography}
\end{document}